# Performance of variable and function selection methods for estimating the non-linear health effects of correlated chemical mixtures: a simulation study

Nina Lazarevic[1,*], Luke D. Knibbs[1], Peter D. Sly[2], Adrian G. Barnett[3]

[1] School of Public Health, Faculty of Medicine, The University of Queensland, Herston QLD 4006, Australia

[2] Children's Health and Environment Program, Child Health Research Centre, The University of Queensland, South Brisbane QLD 4101, Australia

[3] School of Public Health and Social Work, Faculty of Health, Queensland University of Technology, Kelvin Grove QLD 4059, Australia

[*] Corresponding author: n.lazarevic@uq.edu.au

## Abstract

Statistical methods for identifying harmful chemicals in a correlated mixture often assume linearity in exposure–response relationships. Non-monotonic relationships are increasingly recognised (e.g., for endocrine-disrupting chemicals); however, the impact of non-monotonicity on exposure selection has not been evaluated. In a simulation study, we assessed the performance of Bayesian kernel machine regression (BKMR), Bayesian additive regression trees (BART), Bayesian structured additive regression with spike–slab priors (BSTARSS), and lasso penalised regression.

We used data on exposure to 12 phthalates and phenols in pregnant women from the U.S. National Health and Nutrition Examination Survey to simulate realistic exposure data using a multivariate copula. We simulated datasets of size $N = 250$ and compared methods across 32 scenarios, varying by model size and sparsity, signal-to-noise ratio, correlation structure, and exposure–response relationship shapes. We compared methods in terms of their sensitivity, specificity, and estimation accuracy.

In most scenarios, BKMR and BSTARSS achieved moderate to high specificity (0.56–0.91 and 0.57–0.96, respectively) and sensitivity (0.49–0.98 and 0.25–0.97, respectively). BART achieved high specificity ($\geq 0.96$), but low to moderate sensitivity (0.13–0.66). Lasso was highly sensitive (0.75–0.99), except for symmetric inverse-U-shaped relationships ($\leq 0.2$). Performance was affected by the signal-to-noise ratio, but not substantially by the correlation structure.

Penalised regression methods that assume linearity, such as lasso, may not be suitable for studies of environmental chemicals hypothesised to have non-monotonic relationships with outcomes. Instead, BKMR and BSTARSS are attractive methods for flexibly estimating the shapes of exposure–response relationships and selecting among correlated exposures.

# 1 Introduction

Epidemiological studies of environmental exposures are increasingly focusing on the health effects of chemical mixtures, rather than individual chemicals. Mixtures analyses provide better representation of real-world exposure patterns and enable the adjustment of confounding by co-existing toxicants. However, chemical components are often highly correlated (e.g., traffic-related pollutants), which may lead to instability in effect estimates and inflated standard errors in generalised linear models. This can be exacerbated by the inclusion of multiple non-linear terms (e.g., polynomials) when exposure–response relationships are non-linear. These issues present challenges for the statistical identification of important mixture components, i.e., harmful chemicals driving the association between a mixture and health outcome.

Methods for variable selection among multiple correlated exposures have been the subject of recent reviews,[1-6] including penalised regression methods (e.g., lasso and elastic net regression), dimension reduction methods (e.g., sparse partial least squares and supervised principal components analysis), regression tree ensemble methods, and others (e.g., weighted quantile sum regression). These methods produce a sparse solution, which increases interpretability; however, there is no guarantee that the selected exposures are etiologic agents or that the excluded exposures are safe, particularly when exposures are highly correlated.

Recent simulation studies have characterised the sensitivity and rate of false discoveries of variable selection methods in the exposome[7] context (i.e., the totality of environmental exposures throughout the life-course),[8-10] and in assessing interactions between chemical exposures.[6,10] These studies have assumed that exposure–response relationships are linear. However, non-monotonic relationships are biologically plausible and have been observed for metals with dose-dependent effects (e.g., manganese is an essential nutrient at physiologic levels but also an environmental toxicant)[11] and endocrine-disrupting chemicals (EDCs) that mimic the non-monotonic effects of endogenous hormones on endocrine outcomes.[12] The impact of non-monotonicity on the performance of exposure selection methods has not been evaluated.

Standard implementations of many variable selection methods do not accommodate the simultaneous selection of groups of terms associated with an exposure, e.g., spline basis functions. Therefore, methods for variable selection as well as function selection are required for the identification of important mixture components when exposure–response relationships are non-linear. This can be performed, for example, through the use of both sparseness and smoothness penalties in maximum likelihood-based approaches (e.g., the group lasso) and through selection indicators or spike–slab priors in Bayesian models.[13] Alternatively, nonparametric methods may be used to estimate a multivariate exposure–response function.

In this study we compared the performance of three variable selection methods that can model non-linear exposure–response relationships: Bayesian kernel machine regression (BKMR), Bayesian additive regression trees (BART), and Bayesian structured additive regression with spike–slab priors (BSTARSS). We assessed the ability of each method to distinguish outcome-associated exposures and to reveal the shapes of both monotonic and non-monotonic exposure–response relationships when varying the exposure correlation structure. We evaluated whether there was (1) an advantage to the use of these methods when exposure–response relationships are non-linear, over lasso penalised regression assuming linearity, and (2) a cost associated with their use when exposure–response relationships are linear. Building on our previous work,[4] we focused on exposure to EDC-mixtures during pregnancy, as this is a critical period of developmental sensitivity to environmental chemicals.[14-17]

## 2 Methods

### 2.1 NHANES data

We based our simulation study on environmental chemical data from 214 women with positive urinary pregnancy tests in the U.S. National Health and Nutrition Examination Survey (NHANES).[18] We merged data on urinary chemical concentrations measured in the five phthalates and five environmental phenols and parabens surveys between 2005 and 2014 (Table S1 and Table S2, Supplementary Material). Eight phthalate metabolites and four phenols were retained that had fewer than 10% of participants with measurements below the maximum limit of detection (LOD) across surveys (Table S1 and Table S2), and observations below this limit were assigned a value of the maximum LOD/√2. Exposures were corrected for urinary creatinine levels (to account for urine dilution) and expressed in μg/g creatinine, then natural log-transformed and standardized (to keep parameters scale-free and preempt numerical accuracy issues).

### 2.2 Simulating exposure data

We simulated correlated exposure data using a multivariate $t$ copula with truncated kernel-smoothed empirical marginal distributions and the observed Spearman's rank-order correlation structure from the NHANES data. We fitted a selection of multivariate copula types by maximum likelihood and chose the $t$ copula as it had the highest maximum likelihood compared with the Gaussian, Gumbel, Frank, Clayton and Joe copulae. This was performed using the R packages *copula* (version 0.999-18) and *copulaedas* (version 1.4.2).[19,20] The use of a multivariate copula enables the separate specification of marginal distributions and the dependence structure,[20] which allowed us to also simulate a low correlation dataset using half the observed Spearman correlation matrix. The linearity of the dependence structure was first verified through maximal information-based nonparametric exploration (MINE) statistics[21] (Figure S1, Supplementary Material). We assessed fit graphically using scatterplots and nonparametric kernel density estimates of the original and simulated data (Figures S2–S5, Supplemental Material). Correlation heat maps are shown in Figure 1.

### 2.3 Exposure–response relationships and data-generating processes

Our data-generating processes differed by: model size, exposure–response functions $f_j$ ($j = 1, \ldots, J$), the degree of correlation between exposures, and the signal-to-noise ratio. We specified two model sizes of $J = 6$ and $J = 12$ exposure variables $x_{ij}$ ($i = 1, \ldots, N$). A subset of exposures, $J^* = 4$, were assumed to be associated with the outcome; hence, we consider a 'low-sparsity' setting in the $J = 6$ model and a 'high-sparsity' setting in the $J = 12$ model. The response $y_i$ for individual $i$ was generated from the model

$$y_i = \sum_{j=1}^{J^*} f_j(x_{ij}) + e_i,$$

where $e_i \sim N(0, \sigma^2)$.

For simplicity, we assumed no confounding by non-exposure variables, no interaction, and that only phenols were associated with the outcome (i.e., methylparaben (MPB), propylparaben (PPB), benzophenone-3 (BP3), and bisphenol A (BPA)). The eight phthalates were assumed not to be associated with the outcome in the $J = 12$ model, and the two phthalates most correlated with the outcome-associated exposures were included in the $J = 6$ model (i.e., mono-ethyl phthalate and mono-(2-ethyl-5-hydroxyhexyl) phthalate).

We specified four exposure–response functions with two association strengths, shown in Figure 2: linear, non-linear monotonic (S-shaped; using the log-logistic cumulative distribution function),[22] non-monotonic symmetric (inverse-U-shaped; using a quadratic function) and non-monotonic asymmetric (skewed inverse-U-shaped; using a Dawson function).[23] We specified two linear exposure–response association strengths, $f_j(x_{ij}) = \beta_j x_{ij}$ with $\beta_j = 2$ for MPB and BP3 and $\beta_j = 1$ for PPB and BPA. To keep a constant association strength across functions, we scaled each function to have the same area under the curve as the linear function.

We assumed two signal-to-noise ratios, adjusting $\sigma$ so that $R^2$ for the true model corresponded to 10% ('low' signal-to-noise ratio) and 30% ('high' signal-to-noise ratio).

The four exposure–response functions, two levels of exposure correlation, and two signal-to-noise ratios, gave 16 data-generating processes. We simulated datasets of sample size $N = 250$, replicating each data-generating process 100 times to give 1,600 datasets. We estimated models based on the two model size and sparsity settings, producing 32 simulation scenarios.

## 2.4 Statistical methods

We chose three methods for identifying important mixture components and modelling non-linear exposure–response relationships, which are all able to: provide effect estimates with a measure of uncertainty, accommodate several outcome types, model interactions, include linear confounders, and are implemented in accessible software.[4]

### 2.4.1 Bayesian kernel machine regression

BKMR is an approach for mixtures analyses that provides flexible estimation of a multivariate exposure–response function, represented by a Gaussian kernel machine.[24] Expressed as a mixed model and assuming no confounding, BKMR with component-wise variable selection is specified as follows:[24,25]

$$y_i = h_i + e_i, \quad e_i \sim N(0, \sigma^2),$$

$$\mathbf{h} = (h_1, \dots, h_N)' \sim N(\mathbf{0}, \tau \mathbf{K}),$$

where $y_i$ is the response for individual $i$ ($i = 1, \dots, N$), $\mathbf{h}$ is a vector of subject-specific health effects $h_i = h(\mathbf{x}_i)$ with $h(\cdot)$ representing a multivariate exposure–response function, and $\mathbf{x}_i = (x_{i1}, \dots, x_{iJ})$ is the vector of $J$ exposures for subject $i$ ($i = 1, \dots, N$). $\mathbf{K}$ is an $N \times N$ kernel matrix, with $(i, k)$-elements specified by the augmented Gaussian kernel function

$$K(\mathbf{x}_i, \mathbf{x}_k; \mathbf{r}) = \exp\left(-\sum_{j=1}^{J} r_j (x_{ij} - x_{kj})^2\right),$$

where $\mathbf{r} = (r_1, \dots, r_J)'$ is a vector of parameters $r_j$ that control the smoothness of $h(\cdot)$, for which a spike–slab prior is assumed:

$$r_j | \delta_j \overset{\text{prior}}{\sim} (1 - \delta_j) P_0 + \delta_j \Gamma(a_r, b_r),$$

$$\delta_j | \pi \overset{\text{prior}}{\sim} \text{Bernoulli}(\pi),$$

$$\pi \overset{\text{prior}}{\sim} \text{Beta}(a_\pi, b_\pi).$$

$P_0$ is the spike density with point mass at zero and a gamma distribution is specified for the slab component. Here $\delta_j$ are variable selection indicators with prior probability $\pi$. The posterior mean of $\delta_j$ is the posterior inclusion probability of exposure $j$, i.e., a measure of the importance of exposure $j$. The model is estimated by Markov chain Monte Carlo (MCMC), using the Metropolis–Hastings algorithm for $\mathbf{r}$ and $\lambda = \tau\sigma^{-2}$ (a convenient reparameterisation), and a Gibbs sampler for the remaining parameters.[24]

We used the *bkmr* package in R, version 0.2.0, and based our prior specifications on the default implementation.[24] We used a threshold of 0.5 on the posterior inclusion probabilities for variable selection.[26] We assigned $\pi$ a Beta(1,1) prior, so that the prior probability of variable inclusion is 0.5. Prior distributions for $\sigma^{-2}$ and $\lambda$ were assumed to be Gamma with parameters (shape, rate) set to $(a_\sigma, b_\sigma) = (0.001, 0.001)$ and $(a_\lambda, b_\lambda) = (1,1)$, respectively. For the slab component of the prior on $r_j$, we specified a Gamma prior with mean and standard deviation of 0.25, i.e., $(a_r, b_r) = (1,4)$; these values were chosen by fitting frequentist kernel machine regression and observing which values of $r_j$ produced appropriate levels of smoothing.[24] We chose tuning parameters that produced adequate acceptance rates in the Metropolis–Hastings steps (around 20 to 40%); i.e., standard deviations of the gamma proposal distributions for $\lambda$ of 0.5 and 1 for the low and high signal-to-noise ratio datasets, respectively, and for $r_j$ of 0.1 for both the switching and refinement steps (except for the monotonic function and high signal-to-noise ratio datasets, which required a 0.2 standard deviation in the refinement step). We ran the MCMC sampler for 10,000 iterations and discarded the first 8,000 iterations. Convergence diagnostics are presented in Section 2 of the Supplementary Material. We assessed the sensitivity of our results to prior specification in Section 3 of the Supplementary Material.

### 2.4.2 Bayesian additive regression trees

BART is a nonparametric ensemble method, which models an outcome using a sum of regression trees.[27] BART flexibly captures non-linearity and interactions, and imposes no assumptions on the functional forms of exposure–response relationships.[27] BART produces a measure of variable importance by tracking variable inclusion proportions, which enables variable selection with a user-defined threshold.[28] As BART is defined by a Bayesian statistical model, full posterior inference is possible, including exposure effect estimates and credible intervals.[27] The model is

$$y_i = h(\mathbf{x}_i) + e_i, \ e_i \sim N(0, \sigma^2),$$

where $y_i$ is the response and $\mathbf{x}_i = (x_{i1}, \dots, x_{iJ})'$ is a vector of $J$ exposures for individual $i$ ($i = 1, \dots, N$). The multivariate exposure–response function $h(\mathbf{x}_i)$ is approximated by a sum of $K$ regression trees,

$$h(x_i) \approx \sum_{k=1}^{K} g(\mathbf{x}_i; T_k, M_k)$$

Here $T_k$ is the $k^{\text{th}}$ regression tree with terminal node (i.e., "leaf") parameters $M_k = \{\mu_{1k}, \dots, \mu_{n_k k}\}$, for $n_k$ terminal nodes, and the function $g(\cdot)$ assigns $\mu_{lk} \in M_k$ to $\mathbf{x}_i$.[27] A tree $T_k$ consists of non-terminal decision rules (i.e., binary splits of the form $\{x_j \leq c\}$ or $\{x_j > c\}$ for given splitting variables $x_j$ and splitting values $c$), and the set of terminal nodes.[27] Following a sequence of decision rules, each observation is assigned the leaf value $\mu_{lk}$ ($l = 1, \dots, n_k$) of the terminal node. The fitted value $\hat{y}_i = E(y_i|\mathbf{x}_i)$ is then the sum of the $K$ leaf parameters $\mu_{lk}$ assigned to observation $i$.[27]

Individual trees are constrained via a regularisation prior; each tree explains a different small portion of $h(\mathbf{x}_i)$ and the prior ensures that no individual tree is overly influential.[27] The regularisation prior is composed of priors on the tree structure, leaf parameters, and error variance $\sigma^2$, which is assumed independent:[27,29]

$$P\big((T_1, M_1), \ldots, (T_K, M_K), \sigma^2\big) = \left[\sum_k \sum_l P(\mu_{lk}|T_k)P(T_k)\right]P(\sigma^2)$$

$P(T_k)$ takes a form which favours shallow tree structures with fewer splits,[29] with the probability that a depth $d$ node is non-terminal of $\alpha\,(1+d)^{-\beta}$, $\alpha \in (0,1)$ and $\beta \in [0,\infty)$.[27] We used the recommended default values for the hyperparameters $\alpha$ and $\beta$, of $\alpha = 0.95$ and $\beta = 2$, which keeps individual trees small (i.e., greatest probability on trees with 2 or 3 terminal nodes).[27] To complete the specification of $P(T_k)$, a uniform prior is placed on the choice of splitting variable at each node, and a discrete uniform prior is specified for the splitting values.[27]

For $P(\mu_{lk}|T_k)$, a conjugate normal prior is used $\mu_{lk} \sim \text{iid } N(\mu_\mu, \sigma_\mu^2)$,[27] where $\mu_\mu$ is the centre of the response range and $\sigma_\mu^2$ is selected so that the response range centre $\pm\, m = 2$ standard deviations corresponds approximately to 95% coverage of the observed response values.[27] This shrinks the leaf parameters towards the response distribution centre, weakening individual trees.[27]

For $P(\sigma^2)$, a conjugate inverse Gamma prior is used, $\sigma^2 \sim \Gamma^{-1}\left(\frac{\nu}{2}, \frac{\nu\lambda}{2}\right)$.[29] The hyperparameter $\lambda$ was calibrated by first obtaining a data-based estimate $\hat{\sigma}^2$, then setting $\lambda$ such that $P(\sigma < \hat{\sigma}) = q$, i.e., a larger quantile $q$ places more weight on values lower than $\hat{\sigma}$, and setting $\nu$ to obtain an appropriate shape.[27,29] We used recommended values of $(\nu, q) = (3, 0.9)$.[27] We chose $K = 50$ for the number of trees;[29] while larger $K$ have also been recommended, smaller $K$ are preferred for variable selection.[27]

We used the *bartMachine* package in R, version 1.2.4.2.[29] The package allows hyperparameters to be chosen empirically using *k*-fold cross-validation, however this involves a substantial computational burden not feasible in our simulation study. We assessed the sensitivity of our results to prior specification in Section 3 of the Supplementary Material.

BART is estimated by a Bayesian backfitting MCMC algorithm.[27] We set the prior probabilities for proposing grow, prune, and change steps to (0.2, 0.6, 0.2), to achieve adequate acceptance rates. We generated 2,000 draws from the posterior after a burn-in of 4,000 draws. Convergence diagnostics are presented in Section 2 of the Supplementary Material. For variable selection, we chose the local threshold as it gives the least sparse solutions.[28]

*2.4.3 Bayesian structured additive regression with spike–slab priors*

Structured additive regression (STAR) is a flexible modelling framework that allows exposures to be modelled with arbitrary combinations of smooth interactions, random effects, spatial effects, and varying coefficients.[30] Generalised additive models (GAMs) and generalised additive mixed models are special cases of STAR models.[31] BSTARSS extends these models through specification of priors for penalised regression and variable selection, as well as function and model selection; i.e., allowing both individual terms and groups of terms associated with an exposure to be selected or deselected.[30] This enables the simultaneous identification of important exposures, their interactions, and flexible estimation of the shapes of exposure–response relationships. Importantly, BSTARSS differentiates between exposures with no effect, linear effects, and non-linear effects.[30] The model is:[30,32]

$$y_i = h(\eta_i) + e_i, \quad e_i \sim N(0, \sigma^2),$$

$$\sigma^2 \overset{\text{prior}}{\sim} \Gamma^{-1}(a_\sigma, b_\sigma),$$

where $y_i$ is an exponential-family distributed response for individual $i$ ($i = 1, \ldots, N$), $h(\cdot)$ is a known generalised linear model link function, and

$$\eta_i = \eta_0 + X_{iu}\boldsymbol{\beta}_u + \sum_{j=1}^{J} f_j(x_i).$$

Here $\eta_0$ is an optional fixed offset and $X_{iu}\boldsymbol{\beta}_u$ includes terms not subject to selection (such as known linear confounders and a global intercept).

The terms $f_j(x_i)$ can be linear, factors, smooth functions of one or multiple exposures (e.g., splines, tensor products, varying coefficients), random effects, Markov random fields, or interactions between terms.[30] Smooth functions $f_j(x)$ can be represented by a linear combination of $d_j$ basis functions $B_j(\cdot)$, so that

$$f_j(x) = \sum_{k=1}^{d_j} \beta_{jk} B_{jk}(x) = \boldsymbol{B}_j \boldsymbol{\beta}_j,$$

where $B_{jk}(x) = \left(B_{jk}(x_1), \ldots, B_{jk}(x_N)\right)'$ for $j = 1, \ldots, J$.[32] The prior specification assumes that model terms have been reparameterised to separate their penalised and unpenalised parts.[30,32] The coefficient group $\boldsymbol{\beta}_j$, of length $d_j$, is given a parameter-expanded Normal mixture of inverse Gamma distributions prior, denoted by peNMIG($\cdot$):[30]

$$\boldsymbol{\beta}_j = \alpha_j \boldsymbol{\xi}_j \overset{\text{prior}}{\sim} \text{peNMIG}(v_0, w, a_\tau, b_\tau).$$

The prior uses a multiplicative parameter expansion $\boldsymbol{\beta}_j = \alpha_j \boldsymbol{\xi}_j$ that enables simultaneous selection of groups of coefficients, with $\alpha_j$ representing the importance of a coefficient group $\boldsymbol{\beta}_j$ and $\boldsymbol{\xi}_j$ distributing the importance across the elements of $\boldsymbol{\beta}_j$.[30] Each $\xi_{jk}$ is given an iid Normal prior, with a mean of either 1 or –1 with equal probability:[30]

$$\xi_{jk} \mid m_{jk} \overset{\text{prior}}{\sim} \text{iid } N(m_{jk}, 1)$$

$$m_{jk} \overset{\text{prior}}{\sim} \frac{1}{2} I_1(m_{jk}) + \frac{1}{2} I_{-1}(m_{jk})$$

The prior structure for $\alpha_j$ is:[30,32]

$$\alpha_j \mid \gamma_j, \tau_j^2 \overset{\text{prior}}{\sim} N(0, \gamma_j \tau_j^2)$$

$$\gamma_j \mid w \overset{\text{prior}}{\sim} w I_1(\gamma_j) + (1-w) I_{v_0}(\gamma_j)$$

$$\tau_j^2 \overset{\text{prior}}{\sim} \Gamma^{-1}(a_\tau, b_\tau)$$

$$w \overset{\text{prior}}{\sim} \text{Beta}(a_w, b_w)$$

Here $\gamma_j$ is an indicator variable taking the value 1 with probability $w$ and a small value $v_0$ with probability $(1-w)$. The hypervariance $\tau_j^2$ follows an inverse Gamma distribution with shape and rate parameters $(a_\tau, b_\tau)$, $a_\tau \ll b_\tau$.[30] The prior variance $v_j^2 = \gamma_j \tau_j^2$ is a bimodal mixture of inverse Gamma distributions with spike at $\gamma_j = v_0$ (scale $v_0 b_\tau$) and slab at $\gamma_j = 1$ (scale $b_\tau$).[30] The spike part of the prior strongly shrinks coefficients towards zero if $v_0$ is sufficiently small, and its posterior probability gives the probability of exclusion of $\boldsymbol{\beta}_j$ and $f_j(x)$ from the model.[30] The Beta prior on $w$ incorporates prior knowledge on the sparsity of $\boldsymbol{\beta}$.[30]

BSTARSS is implemented in the R package *spikeSlabGAM*, version 1.1-14.[32] We set $v_0 = 0.025$, $(a_\tau, b_\tau) = (5,40)$, $(a_w, b_w) = (1,1)$ (i.e., a uniform prior on $w$), a flat prior on the error variance $(a_\sigma, b_\sigma) = (0.001, 0.001)$, and fit smooth effects using the default reduced-rank representation of 20 cubic B-spline basis functions with equidistant knots and second-order difference penalties.[30] We ran five parallel chains, generating 2,000 draws from each after a burn-in period of 8,000 draws. For variable selection, we used a threshold of 0.5 on the posterior inclusion probability of any term associated with an exposure.[26] Convergence diagnostics are presented in Section 2 of the Supplementary Material. We assessed the sensitivity of our results to prior specification in Section 3 of the Supplementary Material.

*2.4.4 Lasso penalised regression*

Lasso performs variable selection by shrinking the coefficients of exposure variables towards zero. Standard implementations of lasso assume linearity and groups of terms associated with an exposure cannot be selected or dropped simultaneously. When exposure–response relationships are linear, lasso has achieved comparable or superior performance in terms of sensitivity and specificity to competing variable selection methods.[6,9,10] Lasso produces a family of solutions, parameterised by a tuning parameter. Selection of a solution is usually conducted via *k*-fold cross-validation; we used 10-fold cross-validation in the R package *glmnet*, version 2.0-13,[33] and kept the fold assignment constant within each replication of the simulation.

## 2.5 Comparison of methods

We compared methods by the average sensitivity and specificity across the 100 replications, defined as the proportion of outcome-associated and outcome-unassociated exposures, respectively, that were correctly identified. We also measured precision (i.e., the positive predictive value, defined as the proportion of selected exposures that were true positives) and the negative predictive value (defined as the proportion of non-selected exposures that were true negatives). We calculated the $F1$-statistic $= 2 *$ precision $*$ sensitivity/(precision $+$ sensitivity), which reflects the ability of a method to detect outcome-associated exposures while avoiding the selection of unassociated exposures.[34]

To further assess variable selection, we measured the proportion of replications in which all the outcome-associated exposures were ranked higher than the outcome-unassociated exposures (i.e., in terms of the posterior inclusion probabilities for BKMR and BSTARSS, and variable inclusion proportions for BART). Additionally, we measured the mean proportion of outcome-associated exposures that were ranked higher than outcome-unassociated exposures across replications.

To assess estimation accuracy for outcome-associated exposures, we compared the estimated posterior mean (averaged over the post burn-in MCMC samples) evaluated at the $25^{th}$, $50^{th}$, and $75^{th}$ percentiles of each exposure (holding other exposures at their means), to the value of the simulated exposure–response curve using the mean-squared error (defined as the average of the squared differences across 100 replications) and the 90% credible interval coverage (defined as the proportion of times the true value was contained in the 90% credible interval). We compared the ability of each method to estimate the shapes of the exposure–response curves graphically, by plotting posterior means evaluated at every $10^{th}$ percentile of one exposure distribution while holding other exposures at their means. As an oracle (i.e., benchmark) method, we fitted GAMs to the true model of four phenols (R package *mgcv*, version 1.8-23, with restricted maximum likelihood smoothing parameter estimation).[35]

# 3 Results

## 3.1 Variable selection

### 3.1.1 Sensitivity and specificity

Mean sensitivity and specificity for each method and simulation scenario are shown in Figure 3 (and Table S3, Supplementary Material). BKMR and BSTARSS achieved moderate to high sensitivity and specificity for linear and non-monotonic exposure–response relationships across scenarios (BKMR 0.56 to 0.91 sensitivity and 0.49 to 0.98 specificity; BSTARSS 0.57 to 0.96 sensitivity and 0.48 to 0.97 specificity). This was also the case for BKMR when relationships were S-shaped, whereas BSTARSS was less specific for S-shaped relationships (0.25 to 0.81). BART was highly specific but markedly less sensitive than BKMR and BSTARSS (0.13 to 0.66 sensitivity and 0.96 to 1.00 specificity), except in the higher signal-to-noise ratio and sparsity settings.

When exposure–response relationships were linear, lasso achieved high sensitivity (0.86 to 0.99) but lower specificity (0.34 to 0.67) than the Bayesian methods. Lasso was competitive with BSTARSS and BKMR in terms of sensitivity and specificity for S-shaped and asymmetric inverse-U-shaped exposure–response relationships (0.75 to 0.99 sensitivity and 0.34 to 0.69 specificity). However, lasso had very low sensitivity for quadratic relationships (0.13 to 0.20).

The sensitivity and specificity of BART and BSTARSS were robust to changes in the correlation structure, whereas BKMR was more sensitive, but not necessarily more specific, when halving the correlation between chemicals. Lasso tended to be slightly more sensitive but less specific in the lower correlation scenarios.

In almost all cases, increasing the signal-to-noise ratio improved sensitivity. Specificity was also improved for BKMR and in most cases for BSTARSS. However, lasso was less specific in higher signal-to-noise ratio scenarios.

### 3.1.2 Precision and negative predictive value

For BKMR and BSTARSS, we observed high precision (0.74 to 0.97) and moderate to high negative predictive value (0.55 to 0.90) in the low sparsity scenarios. The converse, moderate to high precision (0.50 to 0.97) and high negative predictive value (0.80 to 0.96), was observed in the higher sparsity scenarios (Figure 4 and Table S4, Supplementary Material). This was also the case for lasso when exposure–response relationships were monotonic or asymmetric inverse-U-shaped, although lasso tended to favour negative predictive value to precision, selecting more exposures overall of which fewer were true positives (Table S4). Both precision and negative predictive value were lower for lasso when exposure–response relationships were quadratic. BART performed similarly to BKMR and BSTARSS in the high sparsity scenarios. However, in the low sparsity scenarios, BART was highly precise (0.92 to 1.00) with low negative predictive value (0.36 to 0.47), suggesting that the few exposures selected by BART were likely to be true positives (Table S4).

Overall, for the Bayesian methods, precision and negative predictive value appeared unaffected by the shape of the exposure–response relationships or the exposure correlation structure. The signal-to-noise ratio had a strong positive impact on precision in the high sparsity scenarios (low signal-to-noise ratio scenario precision 0.50 to 0.90 and high signal-to-noise ratio scenario precision 0.70 to 0.98). In the low sparsity scenarios, higher signal-to-noise ratio tended to improve the negative predictive value to a greater degree than the precision.

*3.1.3 F1-statistic*

Considering the F1-statistic (Figure 5 and Table S3, Supplementary Material), a balanced measure of sensitivity and precision, BSTARSS had the best performance of the three Bayesian methods in terms of mean F1 in 25 of 32 scenarios, with comparable or narrower interquartile ranges (IQRs). However, BKMR performed only marginally worse in most low signal-to-noise ratio scenarios and had comparable or better performance in the high signal-to-noise ratio scenarios. BART had the least favourable performance in terms of F1 in the low sparsity scenarios, but had comparable performance to BKMR and BSTARSS in the high sparsity and high signal-to-noise ratio scenarios.

When exposure–response relationships were monotonic or asymmetric inverse-U-shaped, in terms of the mean and interquartile range of F1, lasso tended to perform comparably and sometimes marginally better than BSTARSS and BKMR. However, lasso consistently performed poorly with a low F1 when exposure–response relationships were quadratic.

*3.1.4 Ranking of exposures*

BSTARSS had the highest proportion of replications in which the outcome-associated exposures were all ranked higher than the outcome-unassociated exposures in 30 of 32 scenarios (Figure 6 and Table S5, Supplementary Material), although it was closely followed by BKMR and BART in most cases. This proportion appeared robust to changes in the exposure–response relationship and the exposure correlation structure. However, the proportion was highly affected by the signal-to-noise ratio and sparsity setting: only 2% to 29% of replications yielded the correct ranking across any method in the high sparsity and low signal-to-noise ratio scenarios, whereas 57% to 83% did so in the low sparsity and high signal-to-noise ratio scenarios.

For the mean proportion of outcome-associated exposures that were ranked above outcome-unassociated exposures across replications (Figure 7 and Table S5), most outcome-associated exposures were ranked higher by all methods in the high signal-to-noise ratio scenarios (mean proportions 0.77 to 0.95). In the low signal-to-noise ratio scenarios, BSTARSS achieved the highest mean proportions (0.71 to 0.86 in the low sparsity scenarios and 0.50 to 0.72 in the high sparsity scenarios), although these were closely followed by BKMR (0.64 to 0.80 and 0.42 to 0.61, respectively) and BART (0.58 to 0.69 and 0.43 to 0.53, respectively).

### 3.2 Estimation of exposure–response curves

*3.2.1 Estimation accuracy*

BKMR and BSTARSS achieved similar mean-squared errors to the oracle method, across all scenarios and percentiles of exposure (Figure 8 and Table S6, Supplementary Material). In contrast, BART mean-squared errors were substantially higher than the oracle method, except when exposure–response relationships were asymmetric inverse-U-shaped. Similar conclusions were drawn when considering weak (PPB, BPA) and strong (MPB, BP3) exposure–response relationships separately (Supplementary Material Figure S6 and Figure S7, respectively, and Table S6).

*3.2.2 Credible interval coverage*

BSTARSS and BART both achieved 90% credible interval coverage closely approaching 100% across all scenarios, meaning that the credible intervals were excessively wide (Figure 9 and Table S7, Supplementary Material). BKMR achieved credible interval coverage between 91% and 98% when exposure–response relationships were linear, S-shaped, or quadratic, and between 76% and 89% when relationships were asymmetric inverse-U-shaped. Similar conclusions were drawn when considering weak (PPB, BPA) and strong (MPB, BP3) exposure–response relationships separately (Supplementary Material Figure S8 and Figure S9, respectively, and Table S7).

*3.2.3 Graphical comparison of estimated and true curves*

We present estimated curves for one replication, chosen at random, for MPB (Figure 10). Linear and non-monotonic curves were well estimated, but all methods appeared to have some difficulty with the lower exposure portion of the S-shaped curve. There was also some evidence of under-smoothing by BART. We present all curves estimated by each method for MPB, in Supplemental Material Figures S10–S13.

# 4   Discussion

We assessed the performance of three methods for variable and function selection when exposure–response relationships are non-linear, in a simulation study based on maternal exposure to 12 phthalates and phenols in the NHANES. Our results suggest that BKMR and BSTARSS may be best suited for the analysis of mixtures of correlated chemicals when there is uncertainty regarding the shapes of exposure–response relationships. Both methods performed consistently well across scenarios, balancing moderate to high sensitivity, specificity, precision, and negative predictive value. Moreover, both methods estimated the shapes of exposure–response relationships with error comparable to an oracle method (GAM estimate of the true model). BART had low sensitivity and the highest mean-squared errors of the three Bayesian methods, but was mostly competitive with BKMR and BSTARSS when ranking exposures.

Variable selection requires a choice of threshold for the posterior inclusion probabilities (BSTARSS and BKMR) or variable inclusion proportions (BART). Although the magnitude of the posterior inclusion probabilities may be sensitive to prior and tuning parameter selection,[30,36] our choice of 0.5 (the median probability model)[26] appeared reasonable for BSTARSS and BKMR (on average 4.4 and 3.8 exposures were selected, respectively). For BART, three threshold selection rules have been proposed, of which the local threshold that we used yields the least sparse solutions.[28] However, BART selected too few exposures (1.6 on average) and its sensitivity increased with sparsity, suggesting that this threshold is overly stringent and may be more suitable for sparser problems. In practice, the adequacy of the selected threshold is not known and binary decision making should be avoided in favour of assessing the ranking of exposures according to their posterior inclusion probabilities or variable inclusion proportions.[30]

Our additive (main effects only) data-generating processes favoured BSTARSS, which was specified to model univariate smooths for each exposure. BKMR is, by default, specified to model a multidimensional function for the exposures, and BART is a nonparametric method that imposes no structural assumptions. BKMR and BART do not therefore require *a priori* specification of interactions and are able to automatically identify pair-wise and higher-order interactions. However, if interactions are spurious artefacts in an additive model of univariate non-linear exposure–response functions, BART and BKMR are at risk of interpreting non-linearity as interaction.[37] In contrast, BSTARSS decomposes a multivariate exposure–response function into linear effects, smooth main effects, linear interactions, varying-coefficient terms (linear × smooth interactions), and smooth bivariate interactions.[32] This has the important advantage of allowing the researcher to test linearity and interaction hypotheses; however, it can substantially increase the complexity of the model when there are a large number of exposures.

The use of informative priors in BKMR and BSTARSS (i.e., priors for $r_j$ in BKMR and $\tau_j^2$ in BSTARSS that may influence posterior inference) may have provided these methods an advantage over BART. The variable selection performance of BART may be improved by specifying an informative prior, by giving subsets of exposures greater than equal weight of being selected as splitting variables.[28] Although a correctly specified informative prior can increase power and decrease the chance of a false positive finding,[38] simulations have shown that even incorrectly specified informative priors may benefit variable selection by BART.[28] A common concern is that informative priors are subjective and may excessively influence the posterior;[39] however, sensitivity analyses can assess the impact of varying strengths of prior information on posterior inference (Section 4, Supplementary Material). Moreover, incorporating external knowledge (e.g., from experimental research[40] or meta-analyses) may allow the researcher to explicitly model assumptions that may be a source of bias in conventional (i.e., objective) modelling approaches.[41]

We would expect that linear penalised regression methods have a performance advantage over nonparametric methods when the data-generating process is linear, and this was the case for lasso regression. We also found that lasso variable selection is robust to some degree of non-linearity, specifically, to settings in which the majority of the relationship is approximately linear and the turning point/s occur in the tails of the exposure distribution (i.e., the S-shaped and asymmetric inverse-U-shaped relationships). However, the performance of lasso deteriorated substantially when exposure–response relationships were quadratic, presumably because methods that assume linearity may fit a horizontal line and fail to detect a symmetric U-shaped or inverse-U-shaped relationship. Moreover, we held the shapes of exposure–response relationships constant across outcome-associated exposures. In practice, shapes are likely to vary across chemicals (e.g., a mix of symmetric and asymmetric, monotonic and non-monotonic, relationships), which may affect the ability of lasso to identify and rank non-linear exposures.

While lasso is a sensitive method for identifying outcome-associated exposures when exposure–response relationships are linear or approximately linear, statistical inference is complicated by the highly non-normal finite sample distributions and large sample properties that depend on the choice of tuning parameter.[42] Data-driven approaches such as cross-validation for the selection of tuning parameters may adversely impact variable selection stability.[43] One major advantage of Bayesian penalised regression methods is that inference is based on the marginal posterior of the exposure coefficients, meaning that these methods provide a measure of uncertainty in the coefficient estimates and inference does not depend on the tuning parameters or require asymptotic assumptions.[44] Bayesian methods, therefore, achieve selection stability by allowing parameter and model uncertainty rather than requiring data resampling.[45] Moreover, uncertainty in the tuning parameter estimates can be assessed through their marginal posterior distributions.[44]

Although high correlation between chemicals may complicate the identification of outcome-associated exposures, we observed that the signal-to-noise ratio had a stronger impact on performance than changes to the exposure correlation structure. Penalised regression methods are robust to the effects of collinearity but no method may be able to discriminate between very highly correlated exposures. BKMR may be specified to perform hierarchical variable selection (i.e., to estimate a joint posterior inclusion probability for a group of correlated exposures and conditional probabilities for each exposure),[24] enabling selection of one exposure within the group. However, this can adversely impact stability as the selected exposure may vary with repeated sampling,[46] and may require consultation with subject-matter experts to choose a subset of exposures prior to statistical modelling. Concurvity (i.e., non-linear dependence between exposures) may also adversely impact variable selection performance.[30] This was not the case with the NHANES data but is an important consideration that should be assessed in studies of chemical mixtures (e.g., using MINE statistics).[21]

All of the Bayesian methods we considered provide effect estimates with credible intervals, can adjust for linear confounders that are not subject to selection, and are implemented in accessible software. Several other considerations are important when selecting a method. All three methods can accommodate both continuous (Gaussian) and binary outcomes (BKMR and BART use a Probit model,[29,47] while BSTARSS uses a Binomial model),[32] BSTARSS can additionally accommodate count outcomes (Poisson model),[32] and BKMR and BSTARSS can incorporate random subject-specific intercepts.[32,47] The R packages for both BKMR and BSTARSS require complete datasets, so an additional missing value method may be required such as multiple imputation by chained equations.[48] In contrast, BART automatically handles missing values without imputation, through an extension of the partitioning mechanisms native to tree-based methods.[29,49]

Bayesian methods require careful specification and models should therefore be developed in consultation with subject-matter experts who understand the exposures and their potential health effects. We recommend that researchers fully report their assumptions (including priors and tuning parameters) and use standard model checking procedures. In addition, we recommend that researchers select one method *a priori*, as this increases robustness and reduces the chance of a false positive finding compared with using many methods and publishing the most appealing results.

### 4.1 Limitations

We limited our study to lower dimensional models of 6 and 12 exposures and no confounding by non-exposure variables; however, all of the methods considered are applicable to higher-dimensional settings and allow the inclusion of variables not subject to selection. We did not include interactions between exposures in our data-generating processes. When analysing correlated chemical mixtures, failure to consider non-linearity may lead to biased and false positive interaction effects, and there may be ambiguity in the magnitude of interaction and non-linear main effects when both are included in a model.[37,50] Assessing the performance of methods for detecting non-linear interactions is therefore an important area of future research. Finally, we focused on three variable selection methods that can model non-linear exposure–response relationships, but other methods are available (e.g., variable selection in GAMs[51] and multivariate adaptive regression splines[52]) and have been reviewed elsewhere.[4,13]

### 4.2 Conclusions

We assessed the performance of three Bayesian methods for identifying outcome-associated exposures in a correlated mixture, while varying the shapes of exposure–response relationships. We used a multivariate copula to simulate realistic exposure data based on prenatal exposure to 12 phthalates and phenols in the NHANES. In terms of variable selection performance, there was little cost to the use of BKMR or BSTARSS over lasso penalised regression when exposure–response relationships were linear, and a distinct advantage to their use when exposure–response relationships were non-linear. While Bayesian variable selection methods may require more thoughtful application than their frequentist counterparts, they are able to estimate the shapes of exposure–response relationships, estimate effects at specified exposure levels with credible intervals, and can incorporate external information from experimental studies or meta-analyses. Although mechanisms for non-monotonic EDC dose–response curves in cell-, tissue-, and animal-experimental studies are well understood,[12] and it is widely accepted that non-monotonic relationships can occur in epidemiological studies,[12,53] it is not yet known which shapes are likely to apply to specific EDCs and epidemiological endpoints.[53] Our findings may inform method choice in studies seeking to identify harmful chemicals in a mixture, and to determine the nature of exposure–response relationships.

### Acknowledgements

NL is supported by an Australian Government Research Training Program Scholarship. LDK acknowledges a National Health and Medical Research Council Project Grant (APP1142222). PDS is a Senior Principal Research Fellow of the National Health and Medical Research Council, Australia. AGB is supported by an Australian National Health and Medical Research Council Senior Research Fellowship (APP1117784).

### Competing interests

The authors declare that they have no competing interests.

### Ethics review

The study was deemed to be exempt from ethics review under the Australian National Statement on Ethical Conduct in Human Research and The University of Queensland policy (Clearance Number: 2017001605).

### Data availability statement

The NHANES data are openly available at https://www.cdc.gov/nchs/nhanes/index.htm.[18]

### Reproducibility

The R code to reproduce this simulation study is available at https://github.com/n-lazarevic/.

Figure 1: **Correlation heat maps for (A1) the observed NHANES data, and (A2) all simulated exposure data (using the observed correlation structure), on the top row. The bottom row shows (B1) half the observed NHANES correlation matrix, and (B2) all simulated low correlation exposure data. BPA: bisphenol A; BP3: benzophenone-3; PPB: propylparaben; MPB: methylparaben; MEP: mono-ethyl phthalate; MHH: mono-(2-ethyl-5-hydroxyhexyl) phthalate; MOH: mono-(2-ethyl-5-oxohexyl) phthalate; ECP: mono-2-ethyl-5-carboxypentyl phthalate; COP: mono(carboxyoctyl) phthalate; MZP: mono-benzyl phthalate; MBP: mono-n-butyl phthalate; MIB: mono-isobutyl phthalate.**

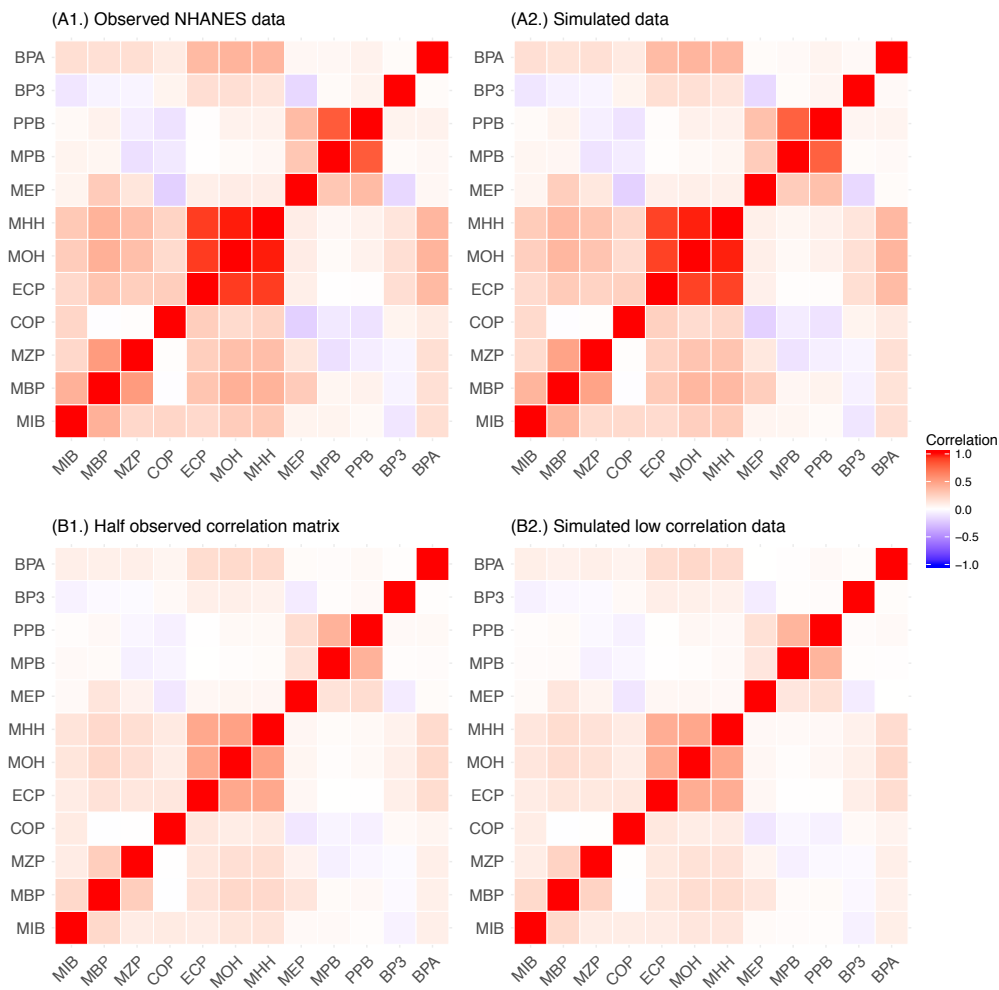

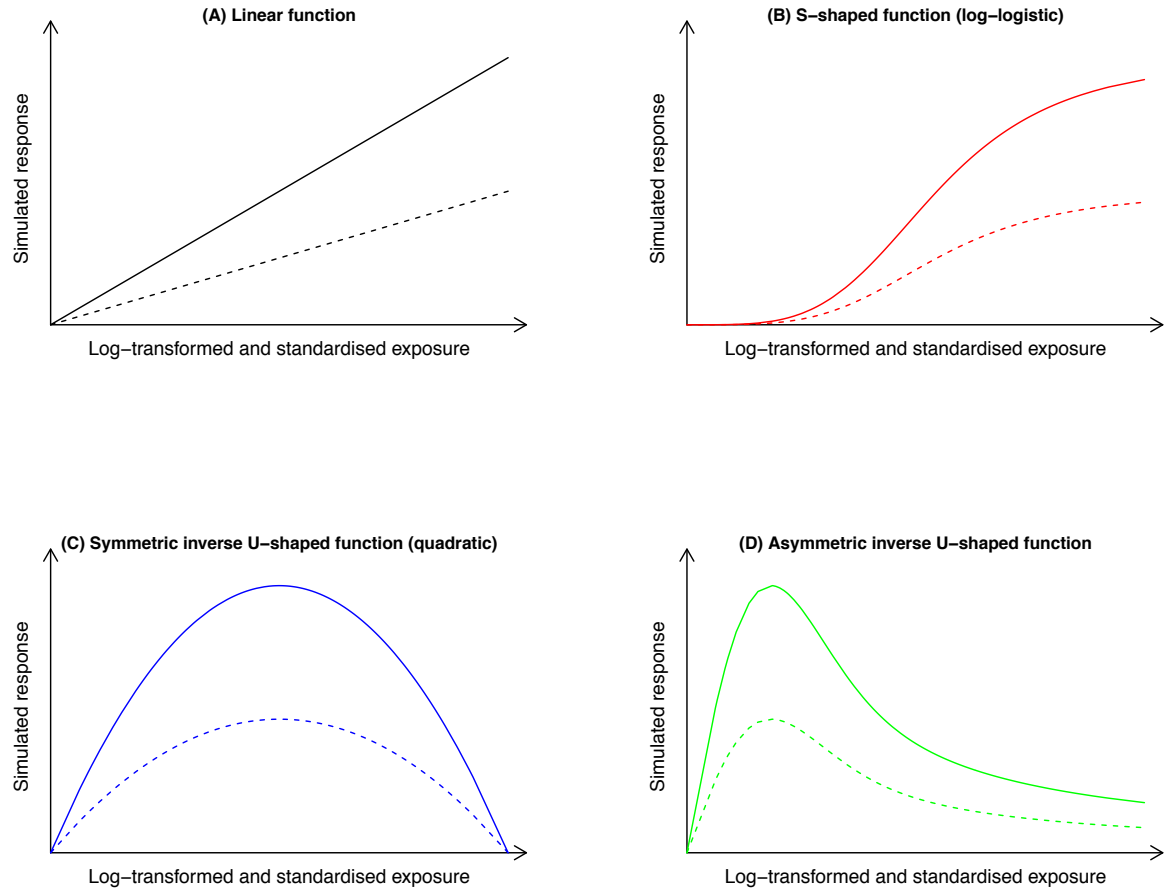

Figure 2: Exposure–response functions with two assumed association strengths; solid line for strong and dashed line for weak.

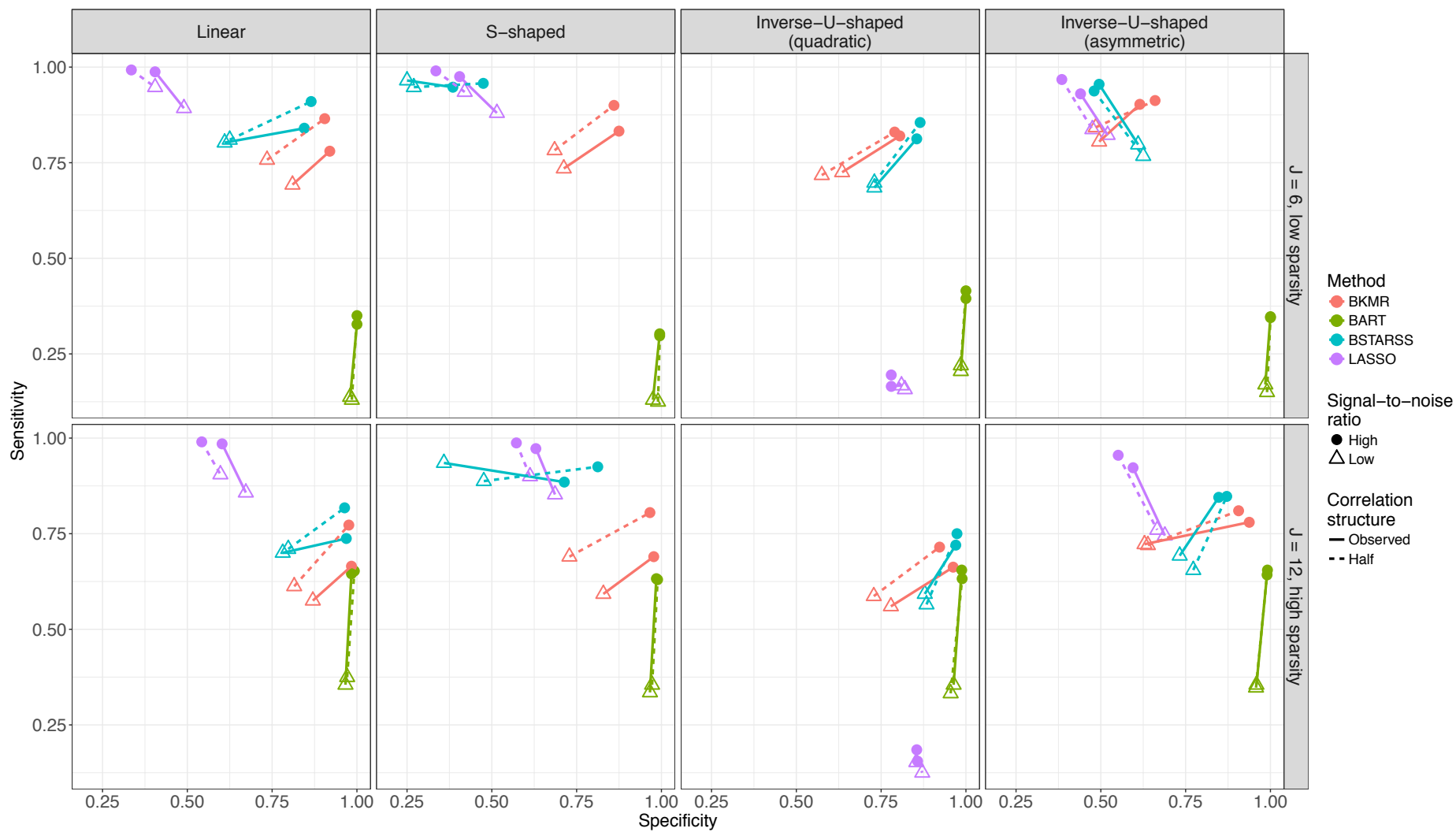

Figure 3: Mean sensitivity and specificity by scenario and method.

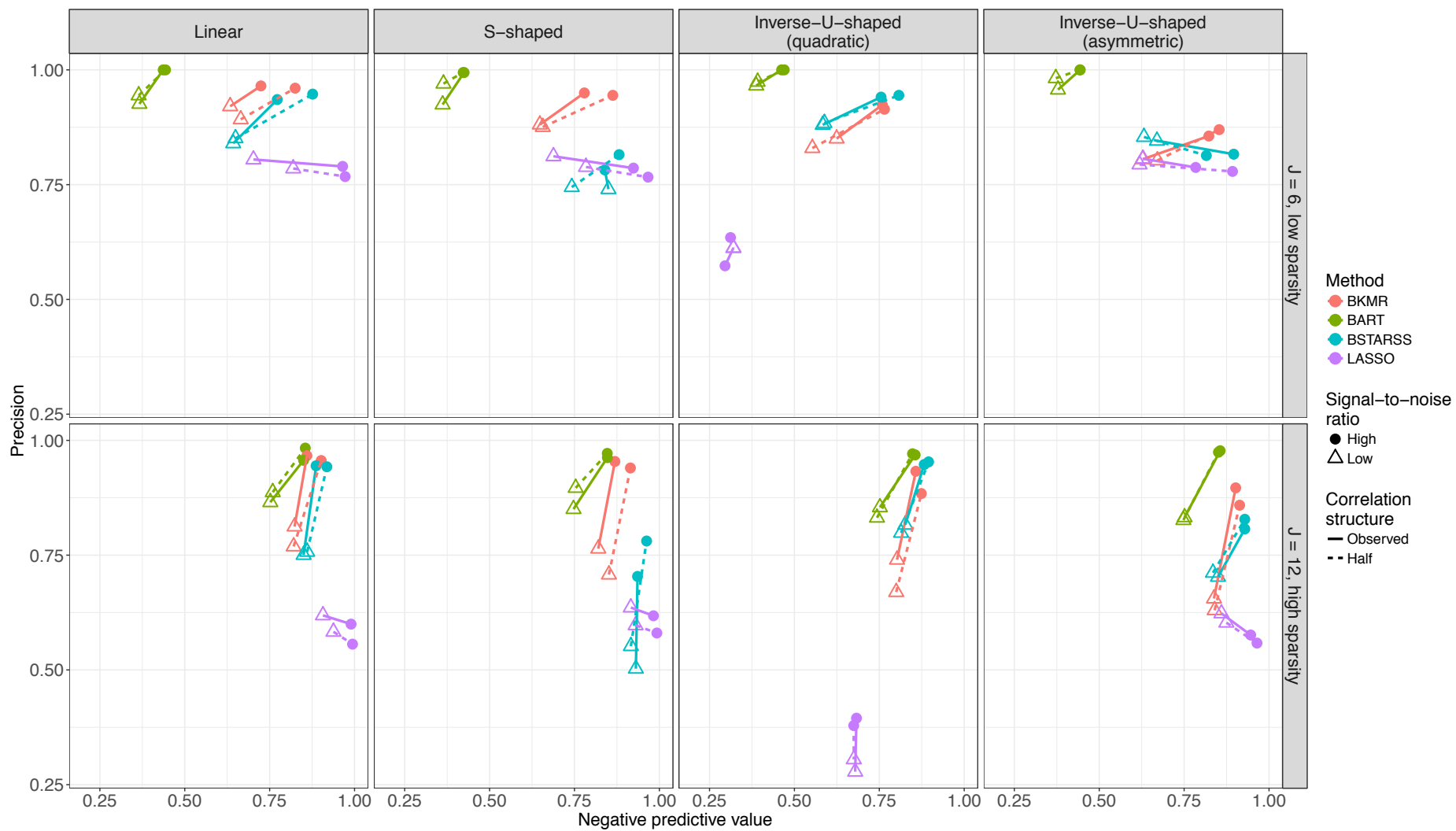

Figure 4: Mean precision and negative predictive value by scenario and method.

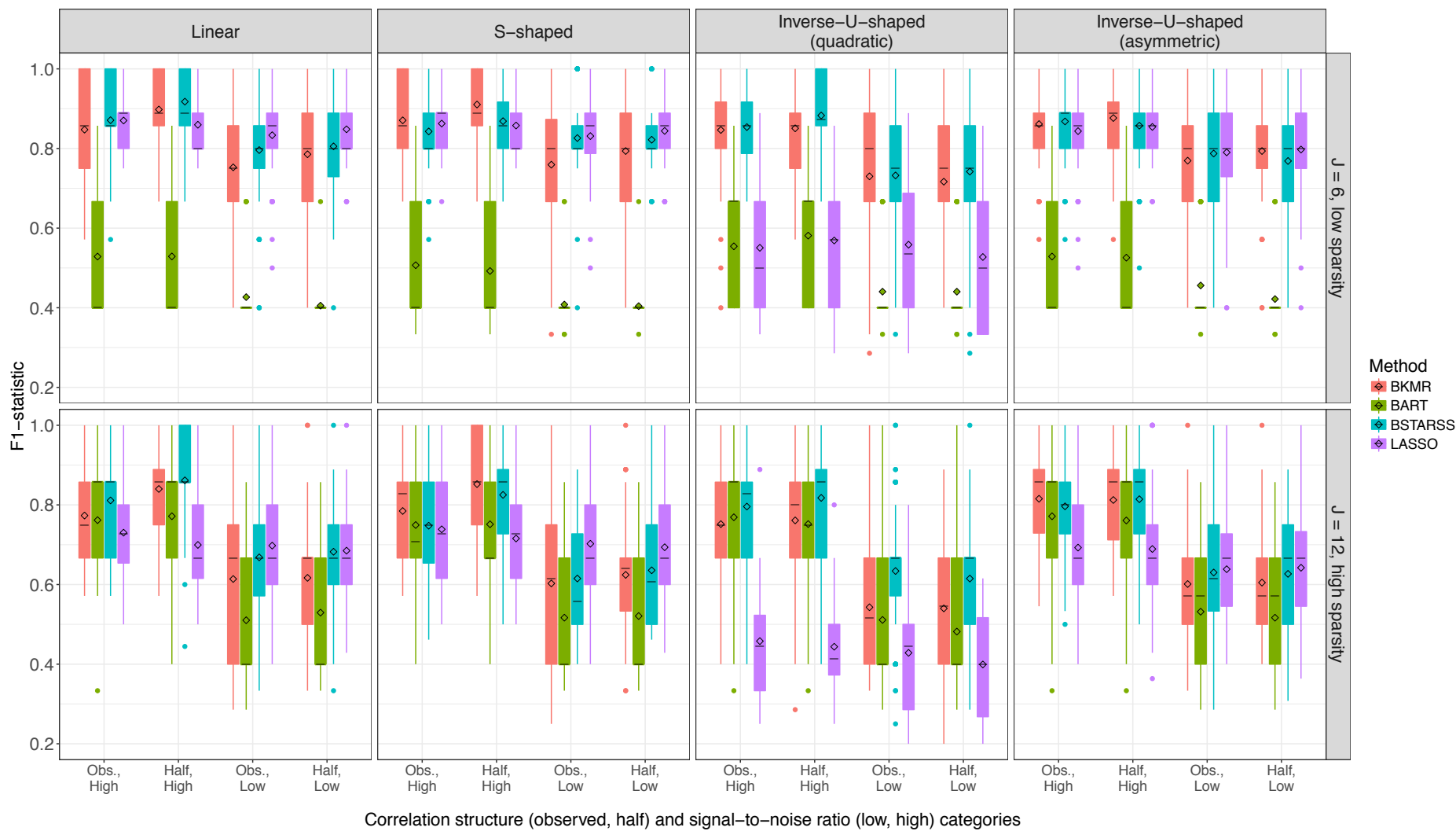

Figure 5: Boxplots of the F1-statistic by scenario and method. Boxplots show the median value (dash) and IQR, with whiskers at ±1.5*IQR. Mean values denoted by diamonds.

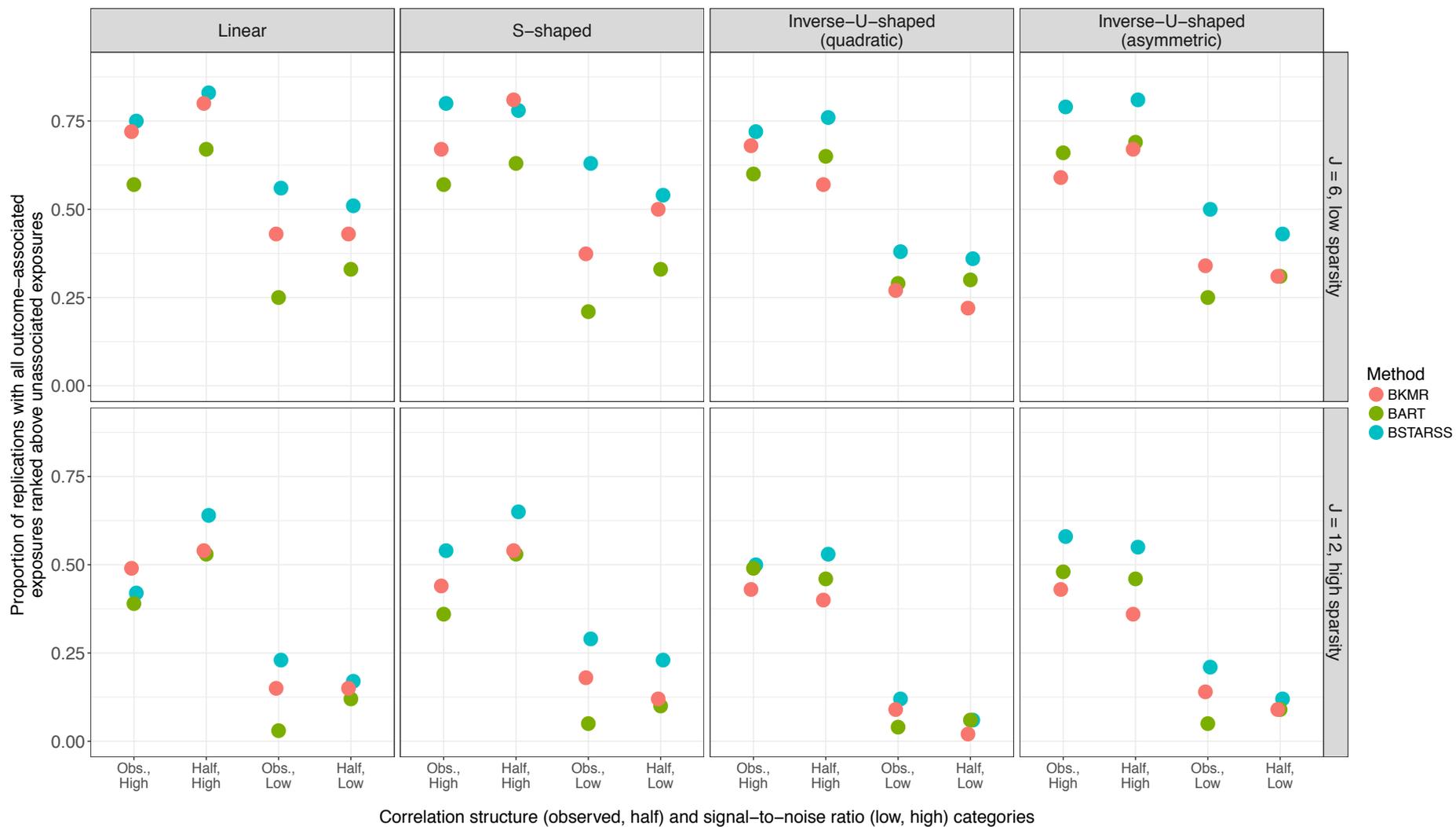

Figure 6: Proportion of replications in which all outcome-associated exposures were ranked above outcome-unassociated exposures (in terms of the posterior inclusion probabilities for BKMR and BSTARSS, and the variable inclusion proportions for BART), by scenario.

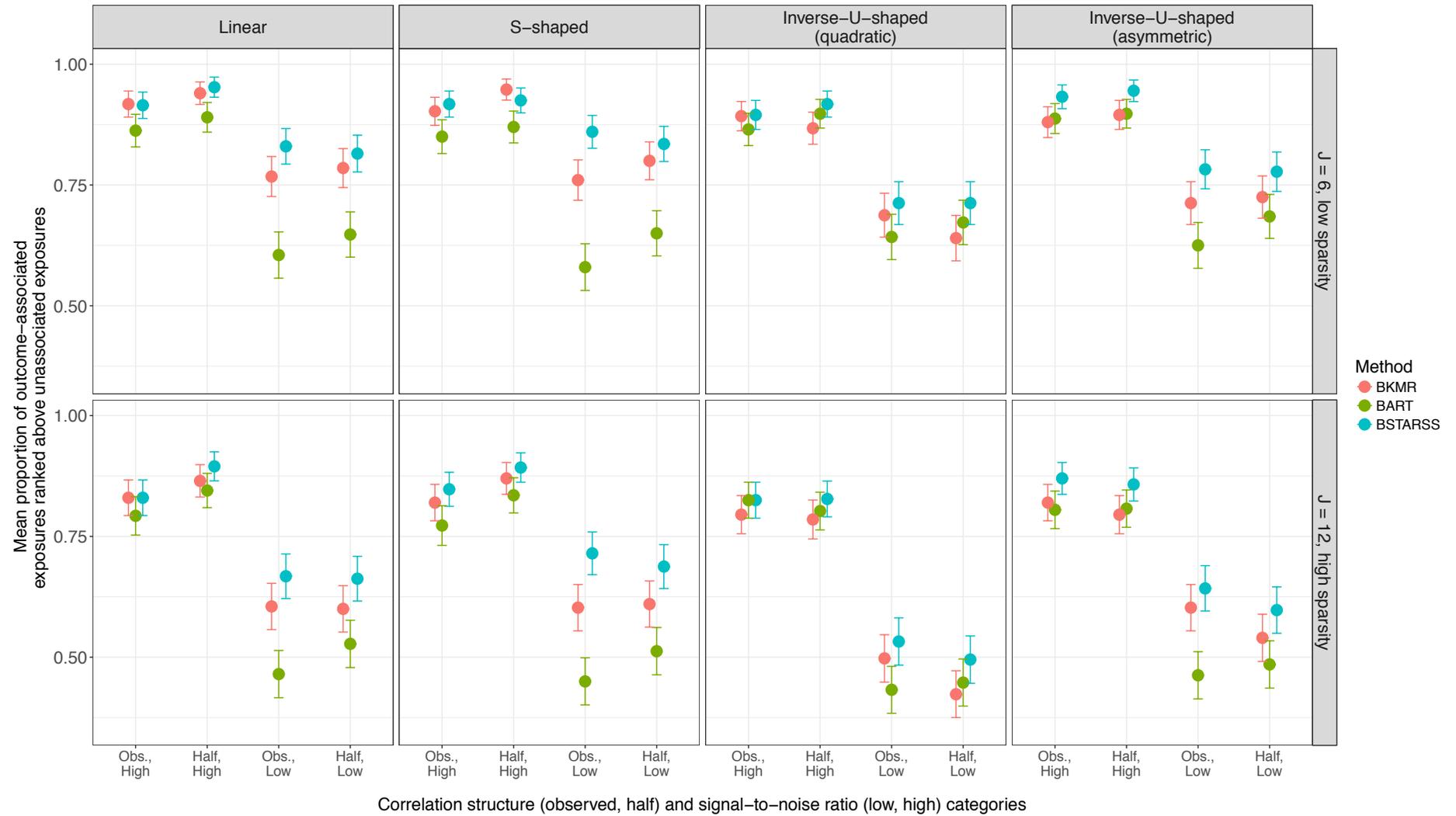

Figure 7: Mean proportion of outcome-associated exposures that were ranked above outcome-unassociated exposures (in terms of the posterior inclusion probabilities for BKMR and BSTARSS, and the variable inclusion proportions for BART), by scenario.

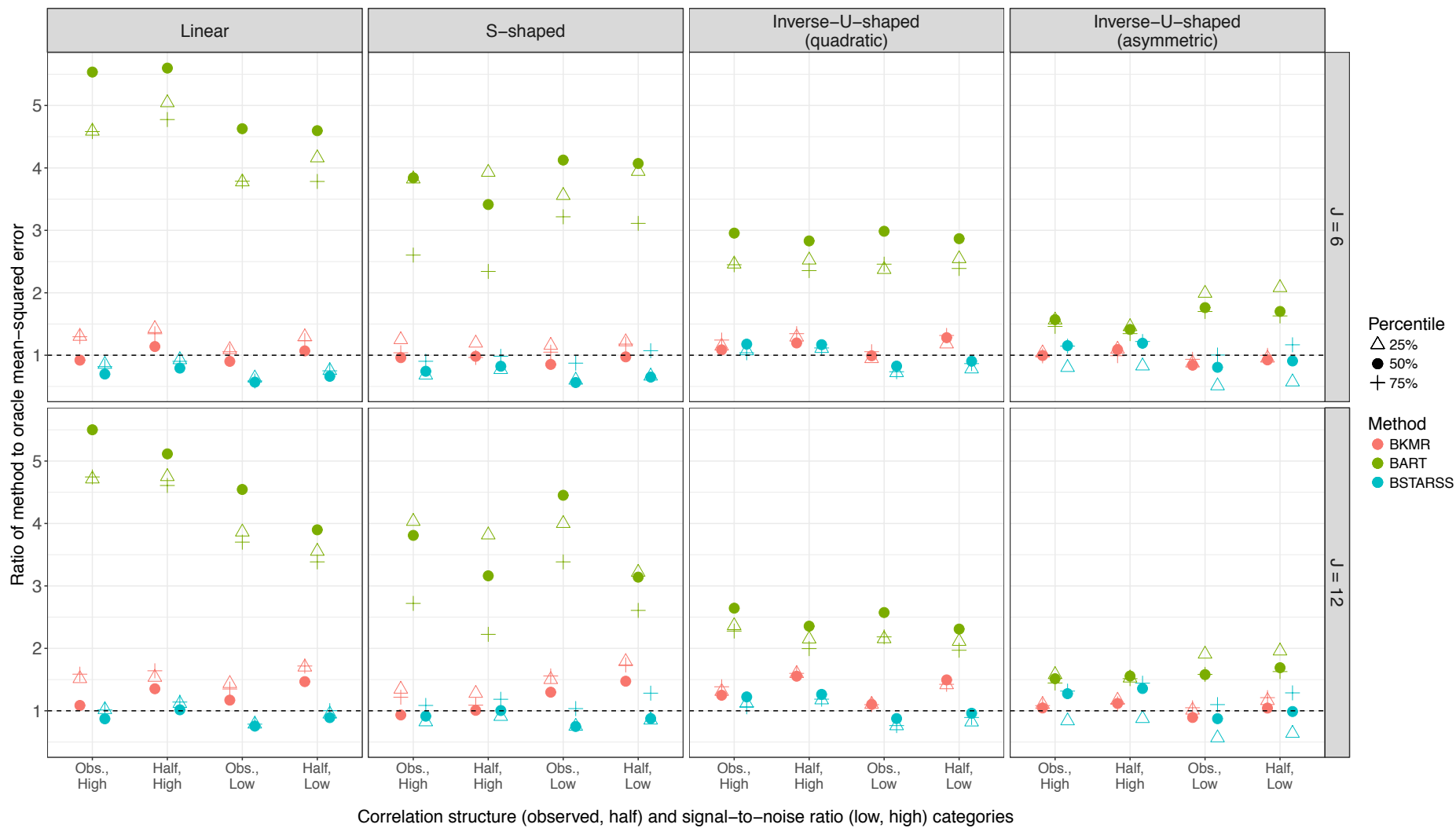

Figure 8: Ratio of method to oracle mean-squared error for outcome-associated exposures, with each exposure evaluated at the 25th, 50th, 75th percentile and other exposures at their mean, by method and scenario. Dashed line at 1 is the targeted ratio.

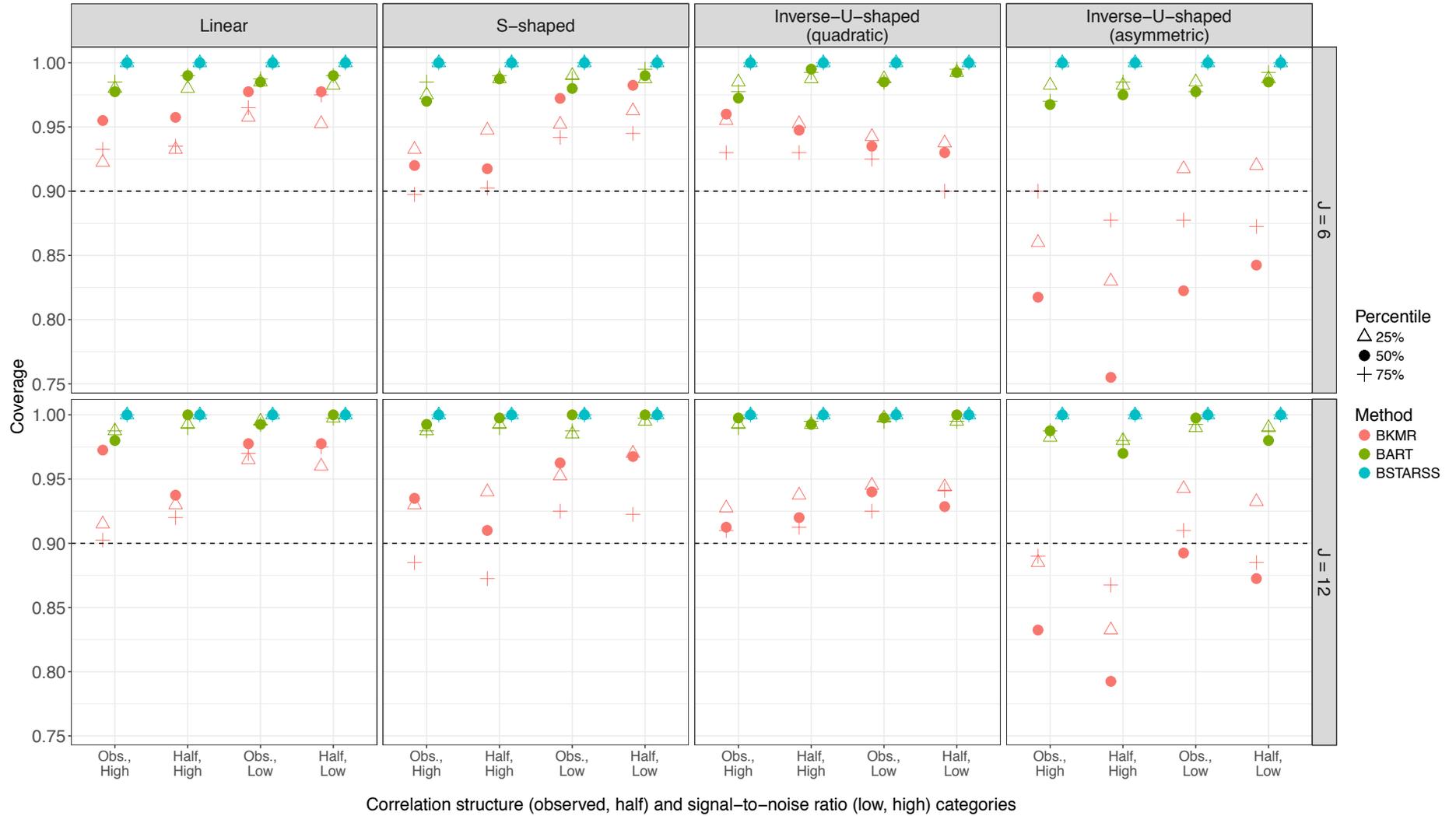

Figure 9: 90% credible interval coverage for outcome-associated exposures, with each exposure evaluated at the 25th, 50th, 75th percentile and other exposures at their mean, by method and scenario. Dashed line at 0.9 is the targeted coverage.

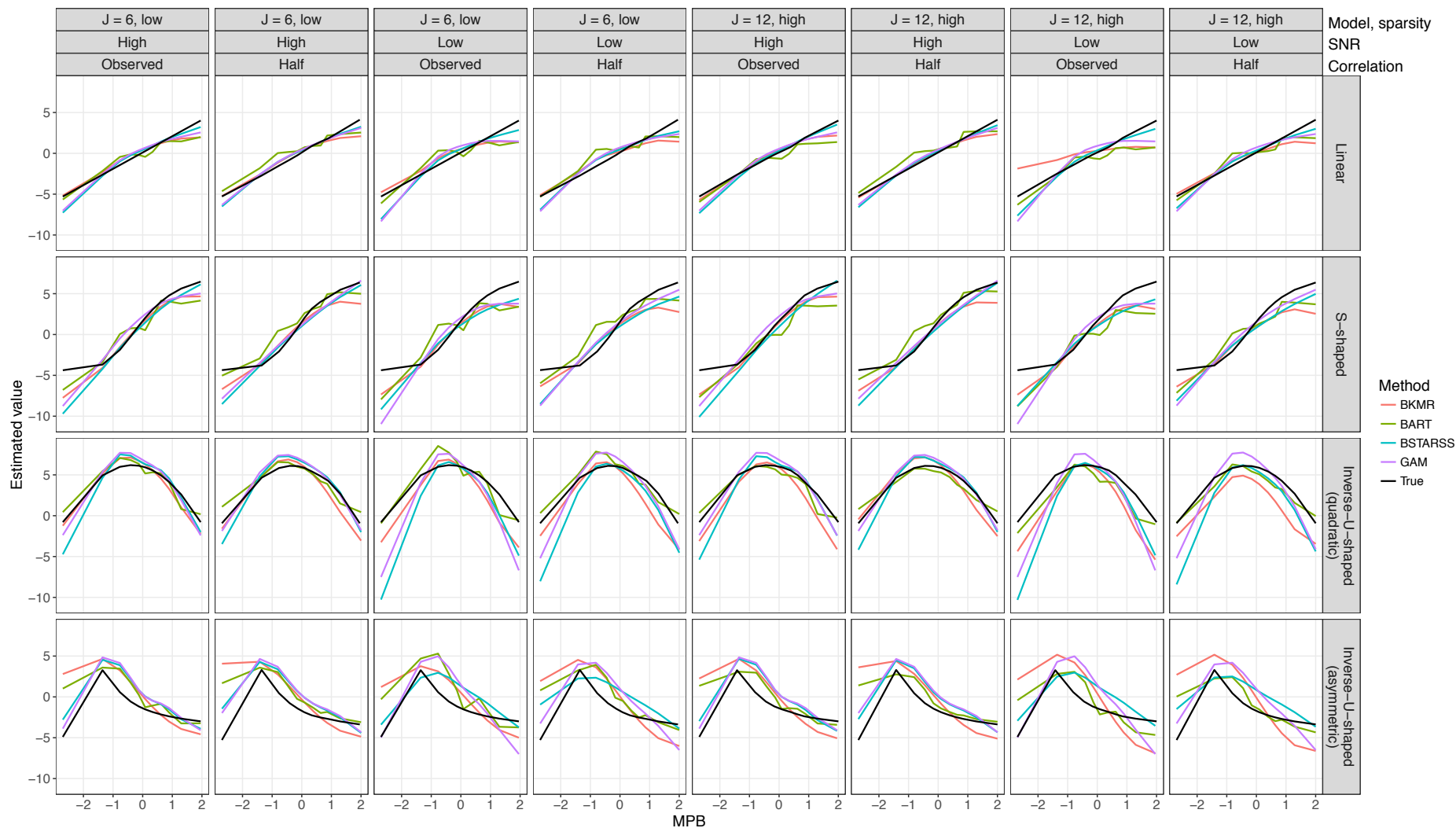

Figure 10: Estimated posterior means evaluated at every tenth percentile of MPB (methylparaben) with other exposures at their means, for each method and scenario in one replication, together with the true curves. SNR: signal-to-noise ratio.